# On the Eclipse of Thales, Cycles and Probabilities

**Miguel Querejeta**

**Abstract.** According to classical tradition, Thales of Miletus predicted the total solar eclipse that took place on 28 May 585 BCE. Even if some authors have flatly denied the possibility of such a prediction, others have struggled to find cycles which would justify the achievement of the philosopher. Some of the proposed cycles have already been refuted, but two of them, namely those of Willy Hartner and Dirk Couprie, remain unchallenged. This paper presents some important objections to these two possibilities, based on the fact that these authors do not list all the eclipses potentially visible by their criteria. In addition, any explanation based on cycles will need to face the complex problem of visibility (smallest observable eclipse, weather…). The present article also includes a statistical study on the predictability of solar eclipses for a variety of periods, similar to that performed by Willy Hartner for lunar eclipses, resulting in lower probabilities in the solar case (and percentages depend on the chosen magnitude limit). The conclusion is that none of the cycles proposed so far provides a satisfactory explanation of the prediction, and, on statistical grounds, none of the periods studied leads to a significant probability of success with solar eclipse cycles.

**Introduction**
From supernovae to comets, from planetary motions to eclipses, the observation of heavenly changes has had a profound influence on humans from the earliest civilizations to the present day. Total solar eclipses stand out among these as some of the most spectacular astronomical events that can be witnessed, and a number of such events have had a significant effect on history. According to Herodotus, Thales of Miletus foretold the loss of daylight which put an end to the battle between the Lydians and Medes; at that time, the battlefield suddenly being plunged into night was regarded as a bad omen:

> They were still warring with equal success, when it happened, at an encounter which occurred in the sixth year, that during the battle the day was suddenly turned to night. Thales of Miletus had foretold this loss of daylight to the Ionians, fixing it within the year in which the change did indeed happen. So when the Lydians and Medes saw the day turned to night, they stopped fighting, and both were the more eager to make peace.[1]

There is some controversy about the translation of the Greek text, and a few editions imply that Thales predicted not only the year, but the date of the eclipse; the latter would make much more sense from the astronomical point of view, as, 'if one can predict an eclipse at all, one can predict it to the day'.[2] Therefore, we will refer to the prediction of the date of the eclipse throughout this paper, and not only to the announcement of the year when it happened. Cicero, Pliny, and Diogenes Laërtius, among others, have also referred to the prediction of Thales, and, in particular, Pliny points out that the event took place in 'the fourth year of the 48 Olympiad',[3] which is widely accepted to correspond to 585/4 BCE. In this context, modern calculations show that a solar eclipse was visible from Asia Minor on 28 May 585 BCE, and the totality path crossed, in all likeliness, the region where the Lydians and Medes were fighting.[4] Yet, could the Milesian philosopher have predicted such a special occurrence?

Strenuous efforts have been made to either support or deny the prediction of Thales. Since the nineteenth century, when the first serious attempts were made to pinpoint the moment of the famous eclipse,[5] an integral part of the debate has depended on whether calculations yielded totality over Asia Minor for each of the possible dates; after all, the words of Herodotus 'the day was suddenly turned to night' would make little sense had the eclipse been only partial over the battlefield. Even quite recently, authors have tried to justify the fact that the eclipse of Thales should correspond to a variety of dates.[6] However, it has been made clear by Stephenson and Fatoohi that only the eclipse of 28 May 585 BCE could match the visibility required from Asia Minor.[7]

---

1 Herodotus, *Histories,* I.74.2–3, trans. R. Godley (Cambridge Mass.: Harvard University Press, 1920).

2 Dimitri Panchenko, 'Thales's Prediction of a Solar Eclipse', *Journal for the History of Astronomy*, Vol. 25 (1994): p. 275.

3 Pliny the Elder, *Naturalis Historia*, II.53, trans. H. Rackham (Cambridge Mass.: Harvard University Press, 1942).

4 All eclipse computations correspond to Fred Espenak and Chris O'Byrne, NASA Goddard Space Flight Centre; this will be commented in detail later. The battle took place somewhere in the Anatolia peninsula, which was almost completely covered by the totality path.

The complexity of the problem also stems from the fact that the choice of the date of the eclipse has strong historiographic implications, as intricate chronological issues are involved.[8] This problem is to some extent independent from the debate as to whether Thales could have predicted the eclipse: we now know for sure that there was a total solar eclipse in Asia Minor on 28 May 585 BCE, and this date is in accordance with the one given by Pliny. The question of the prediction, on the contrary, is extremely controversial. Some scholars, including Martin and Neugebauer,[9] have flatly denied the prediction, while others have struggled to find a numerical cycle by means of which the prediction could have been carried out. Most of these conjectures have already been refuted,[10] but there are two of them which stand unchallenged for the moment: the mechanism proposed by Willy Hartner,[11] and a 'false cycle' suggested more recently by Dirk Couprie.[12]

---

5 The first significant attempts correspond to Sir Francis Baily (1811), Sir George B. Airy (1853), and Prof Simon Newcomb (1878). See John Stockwell, 'On the eclipse predicted by Thales', *Popular Astronomy*, Vol. 9 (1901): pp. 376–89.

6 The eclipse of 30 September 610 BCE has traditionally been considered instead of the one in 585 BCE in terms of having some chronological advantages. Panchenko points to the eclipses on 21 September 582 BCE and 16 March 581 BCE based on the assumption that Thales should have used the *Exeligmos* to predict the eclipse. See Panchenko, 'Thales', pp. 275–88.

7 F. R. Stephenson and L. J. Fatoohi, 'Thales's Prediction of a Solar Eclipse', *Journal for the History of Astronomy*, Vol. 28 (1997): pp. 279–82.

8 See, for example, A. A. Mosshammer, 'Thales' Eclipse', *Transactions of the American Philological Association*, Vol. 111 (1981): pp. 145–55.

9 Thomas-Henri Martin, 'Sur quelques prédictions d'éclipses mentionnées par des auteurs anciens', *Révue Archéologique*, Vol. 9 (1864): pp. 170–99. Otto Neugebauer, *The Exact Sciences in Antiquity*, 2nd ed. (New York: Dover Publications, 1969).

The aim of this paper is to assess the validity of these two proposals, and also to perform a statistical study analogous to the one carried out by Willy Hartner for lunar eclipses, in order to know whether the same results apply to the solar case. This will permit us to conclude whether it is reasonable to defend that Thales made his prediction based on one of the cycles which have been discussed in the literature so far.

**The Hypothesis of Dirk Couprie**
One of the two cycles that remain uncontested corresponds to a recent study by Dirk Couprie.[13] After refuting the proposal of Patricia O'Grady,[14] in his attempt to justify Herodotus' claim, Couprie resorts to a 'false prediction' by Thales. In his paper, the author suggests that Thales could have deduced that solar eclipses come in clusters of three events, judging by the data presumably available to the Milesian philosopher. The pattern, consisting of three consecutive eclipses, would span a total of 35 lunations (the time between eclipses in a cluster being either 17–18 or 18–17 lunations), and these clusters would be separated by much larger gaps (see Table 1).

---

10 For example, the eclipse prediction suggested in Panchenko, 'Thales', was proved to be extremely unlikely by Stephenson & Fatoohi, 'Thales's Prediction'; similarly, the prediction mechanism proposed by Patricia O'Grady has recently been refuted by Dirk Couprie (see note 14).

11 Willy Hartner, 'Eclipse Periods and Thales' Prediction of a Solar Eclipse: Historic Truth and Modern Myth', *Centaurus* Vol. 14 (1969): pp. 60–71.

12 Dirk L. Couprie, 'How Thales Was Able to "Predict" a Solar Eclipse without the Help of Alleged Mesopotamian Wisdom', *Early Science and Medicine*, Vol. 9 (2004): pp. 321–37.

13 Couprie, 'Thales'.

14 Patricia O'Grady, *Thales of Miletus. The Beginnings of Western Science and Philosophy* (Aldershot: Ashgate, 2002). She suggests that Thales used a cycle of 23 ½ months, but this method yields a much smaller probability of success (23%) than initially calculated by O'Grady.

One of the most evident problems that this hypothesis poses is the need for a long list of eclipse records, as the author himself acknowledges. In particular, the author sets an arbitrary limit of magnitude 0.5 to the smallest observable eclipse.[15] However, even more important than this is the problem of cloudiness, since an important fraction of the eclipses making up the list may well have gone unnoticed due to unfavourable weather conditions. Still, the possibility of all those eclipses being recorded exists in principle, so the visibility issue cannot be regarded as a conclusive counter-argument; there is an additional objection that can be raised to this cycle, however, based on the fact that the eclipse list is not exhaustive according to the criterium chosen by the scholar.

The author assumes a limit in magnitude of 0.5 to the eclipses that Thales could have observed, 'unless he had a special reason to expect one',[16] which justifies the inclusion of the eclipse of 9 May 594 BCE (magnitude 0.46). However, Couprie consciously ignores the eclipse of 17 April 611 BCE, which had a magnitude of 0.45 *at the moment of sunset*[17]; this would have permitted an individual to see virtually half of the solar diameter eclipsed and, due to the proximity to the horizon, it would have been possible to observe it with the unaided eye. Moreover, Couprie admits records since 635 BCE, but makes no mention of the eclipse of 17 June 633 BCE, which attained a maximum magnitude of 0.66 shortly before sunset (therefore, weather permitting, probably observable). Similarly, the eclipse of 18 May 603 BCE (magnitude 0.50) is missing from his list. If we include all these eclipses in the record that Thales could have made himself, the 'obviousness of the pattern of clusters' seems to vanish (see Table 2).[18]

---

15 Magnitude is defined as the fraction of the Sun's diameter obscured by the Moon, expressed here in decimal units; it is sometimes expressed in digits, or even as the area of the sun that is blocked by the lunar disk (obscuration). In any case, increasing magnitude in terms of diameter (decimal or digits) also implies an increase in the obscuration, so they could ultimately be seen as different measures of the same thing.

16 Couprie, 'Thales', p. 331.

17 All magnitudes and local circumstances of eclipses according to NASA data (Fred Espenak and Chris O'Byrne, NASA's GSFC).

18 Couprie, 'Thales', p. 337.

**The Hypothesis of Willy Hartner**
One of the most ambitious and interesting studies on the eclipse of Thales was undertaken in 1969 by Willy Hartner.[19] The first part of his article is devoted to a statistical survey of lunar eclipses and their predictability according to a number of cycles. The second part lists 29 solar eclipses which might have been observable from Miletus, and based on the analysis of the repetition pattern, Hartner proposes that Thales intended to predict the eclipse of 18 May 584 BCE. Such a prediction was made in some vague terms due to calendrical difficulties, argues Hartner, which would explain why the Milesian philosopher was acclaimed when a total solar eclipse took place one year earlier than he expected.

The objections made to this hypothesis so far include the requirement of a long and precise record of eclipses, as mentioned in the previous section, as well as the lack of security in the prediction due to the overwhelming number of ongoing cycles. In addition to this, it is important to note that, in his detailed study, Hartner does not list *all the solar eclipses* visible from Miletus in the time-period considered. One would naturally think that he just sets a limit to what eclipses could have been observed by considering their visibility magnitude. However, if we use the excellent ephemerides now available from NASA, we can see that Hartner's 'Table 3' includes eclipses of magnitude down to 0.32 (10 November 687, 28 June 596, and 18 May 584 BCE all have magnitude 0.32), but ignores six eclipses of magnitude greater than 0.5[20]: for example, the eclipse on 17 June 633 BCE reached a maximum magnitude of 0.66 just before sunset; thus, weather permitting, it may well have been observable. The inclusion of these eclipses in the list completely alters the 'gaps' that make Hartner suggest that Thales could have deduced a certain number of cycles from these data. In spite of this, Hartner's probabilistic approach to lunar eclipse prediction remains of considerable interest, and it would be valuable to perform a similar statistical study for solar eclipses, taking their visibility into account.

---

19 Hartner, 'Eclipse Periods'.

20 Eclipses of magnitude more than 0.5 omitted by Hartner in the period considered (magnitude in brackets): BCE 17 July 709 (0.67), 5 May 705 (0.54), 5 Mar 702 (0.53), 28 July 691 (0.66), 17 June 633 (0.66), 23 December 596 (0.61). Moreover, the eclipses on BCE 17 April 611 (0.45) and 1 October 583 (0.36), which attained those maximum magnitudes at sunset, could have been easily observable (it is easier to gaze an eclipse of a given magnitude if the maximum occurs near the horizon, because the atmosphere of the Earth blocks most of the light from the Sun and, therefore, even a small eclipse can be noticed with the unaided eye provided that it is not cloudy). The eclipse of 17 June 679 BCE, which reached its maximum (0.48) at a rather low Sun altitude (18º), is also worth mentioning.

**Predictability of Solar Eclipses Using A Variety of Cycles**
In the first part of his article, Willy Hartner calculates the probabilities of a *lunar* eclipse being followed by another one after a certain number of periods (some of them traditionally identified as 'eclipse cycles'), concluding that the *triple Saros* or *Exeligmos* 'deserves the praise unjustly wasted on the *Saros*'.[21] It is important to note that conclusions about the eclipse of Thales have been derived from these results; that is, conclusions on the predictability of lunar eclipses have been extrapolated to the prediction of solar eclipses.[22] Consequently, it would be interesting to undertake a similar study based on a statistical record of the solar eclipses potentially visible from Miletus, analogous to what Hartner did for lunar eclipses. We are going to calculate the percentages of those solar eclipses that repeat after a number of cycles (those listed by Hartner) in a given time-interval. It would also be good to quantify whether the resulting percentages depend on the visibility limit that we impose and, for this purpose, all results will be computed for three different arbitrary magnitude limits (0.25, 0.50, and 0.75).

It is important to note that not all solar eclipses can in reality be observed, as we have already pointed out. On the one hand, the magnitude limit required for an eclipse to be recorded depends strongly on the method used to observe it (naked eye, pinhole effect, reflection…), and also on whether observers are carefully watching at the sun because they expect an eclipse to happen. The latter has usually been assumed to be true in the literature, and it has been explicitly stated by Couprie, who suggests that Thales may have the habit of observing the reflection of the sun on a liquid surface every new moon close to an

---

21 Hartner, 'Eclipse Periods', p. 60.

22 Hartner himself derives conclusions for the solar eclipse of Thales from these lunar probabilities (Hartner, 'Eclipse Periods', p. 65), and so do Panchenko (Panchenko, 'Thales', p. 280) and Couprie (Couprie, 'Thales', p. 322) in their papers.

Eclipse Season.[23] On the other hand, a more crucial matter implies the meteorological conditions necessary for an eclipse to be visible. Even if we set a reasonable (but obviously arbitrary) limit to the minimum observable eclipse magnitude, when we work out percentages as Hartner did, we are at most obtaining an upper limit to the average probability of repetition after each of the periods. This is because *at most* those eclipses will be observed, but in all probability, many of them will pass unnoticed due to cloudiness.

Using a *JavaScript* application available through the NASA website, local circumstances can be retrieved for all eclipses of a given type visible from the chosen location over a certain time-period. This extremely powerful tool uses the best available corrections, so the ephemerides can be considered most reliable.[24] The circumstances of the eclipses potentially observable from Miletus from 1000 BCE to 501 BCE have first been obtained: we are not assuming that Thales had access to eclipse records for such a long period, but this arbitrary length has rather been selected in order to have a sufficiently large number of events to work out statistics. The choice of a different geographical location[25] or an appreciably different historical period could lead to different results, as the orbital properties of the Moon slightly change with time, and, depending on the latitude, a different amount of eclipses occur in a given time interval.

Results are shown in Table 3. Of course, this set of periods does not list all the possible repetition patterns, but it is quite comprehensive, as it contains all the significant multiples of less than 100 years, including famous cycles like the *Eclipse Seasons* (6 lunations), *Saros* (223 lunations) or *Exeligmos* (669 lunations). We find that the *Exeligmos* provides the best results for solar eclipses as well, but even for the lowest magnitude limit considered (0.25) the resulting percentage (58.5%) is smaller than the one found by Hartner for lunar eclipses (76.2%), and it reduces down to 22.7% when we require a minimum magnitude of 0.75. Surprisingly, Hartner's second best rated cycle (T*, 1074 lunations) performs quite poorly for solar eclipses (8.5%, 2.3%, 0.0% for the different magnitude limits, compared to Hartner's 60.6%). It can also be noted that periods marked by Hartner with an asterisk (cycles in which consecutive eclipses occur in opposite nodes) clearly provide worse results in the case of solar eclipses, as opposed to lunar eclipses, where this distinction was not significant.

---

23 Couprie, 'Thales', p. 331.

24 URL: http://eclipse.gsfc.nasa.gov/JSEX/JSEX-index.html [accessed 2 June 2011]. This *JavaScript* calculator uses the same Besselian elements as the *Five Millennium Canon of Solar Eclipses: -1999 to +3000*, and the values for ΔT are calibrated using historical eclipses, based on the work by L. Morrison and F. R. Stephenson, 'Historical Values of the Earth's Clock Error ΔT and the Calculation of Eclipses', *Journal for the History of Astronomy*, Vol. 35 (2004): pp. 327–36.

25 Miletus: 27º20' E, 37º30' N, the same coordinates chosen both by Hartner and Couprie, to avoid divergences for this reason; altitude: 10m over the sea level.

With this statistical study it has been shown that repetition percentages for solar eclipses are always smaller than those obtained for lunar eclipses. Moreover, our percentages do not take meteorological difficulties into account, which are certainly not negligible; therefore, true percentages would in reality be even smaller than the ones listed in Table 3. As already noted in the existing literature, the eclipse preceding 28 May 585 BCE by an *Exeligmos* period was not visible from Miletus, which discounts the possibility of a prediction based on this method. On the contrary, one *Saros* before, a partial solar eclipse was visible from Miletus[26]; judging by the percentages obtained, however, the cycle cannot be considered a reliable method for predicting solar eclipses (6.9% for a magnitude of 0.5), and thus seems unlikely as an explanation for the prediction of Thales. Even if we are not in a position to completely rule out the possibility of a prediction based on cycles, we can state that, if such a cyclic prediction took place, the probabilities of guessing right suggest that it should be seen as a lucky guess.

**Conclusions**

We have first considered the two cyclic approaches that stood unchallenged (all the others had already been refuted). The conclusion is that none of those conjectures can be regarded as serious explanations of the problematic prediction of Thales: in addition to requiring the existence of long and precise eclipse records, which clashes with the meteorological visibility difficulties, both cycles that have been examined overlook a number of eclipses which match the visibility criteria and, consequently, the patterns suggested seem to disappear.

---

26 It is the solar eclipse of 18 May 603 BCE, which reached a maximum magnitude of 0.83 in Miletus.

   Secondly, Hartner proposed a method to calculate repetition probabilities for lunar eclipse cycles, and a similar technique has been applied here to the solar case. The percentages obtained for each of the periods are significantly smaller than those found by Hartner for lunar eclipses, and our results diminish as we require higher magnitude limits. Therefore, another important conclusion is that the probabilistic calculations for lunar eclipses cannot be extrapolated to the prediction of solar eclipses.

   It has also been stressed throughout the paper that visibility plays a key role in any attempt to explain the prediction by means of cycles. In addition to the difficulty of setting a limit to what the smallest observable eclipse is, cloudiness would add a random component that prevents observers from regarding certain eclipses. As a consequence, all the repetition percentages that we have calculated must be understood as an upper limit to the probability of an eclipse repeating after a given period. Therefore, and since the *Exeligmos* must be ruled out because the preceding eclipse was not visible from Miletus, we can conclude that any of the cycles considered implies a small probability of correctly predicting a solar eclipse (in all of them, even for the smallest magnitude limit of 0.25, failure turned out to be more likely than guessing right). This means that, if Thales used a cyclic mechanism at all, his prediction can hardly be considered fully scientific.

| Date of solar eclipse | Maximum phase | Local time at maximum phase | Altitude of sun at maximum phase | Lunations elapsed since last solar eclipse |
|---|---|---|---|---|
| 30 Sept 610 | 0.59 | 8.6 h. | +30° | 317 |
| 13 Feb 608 | 0.76 | 15.2 h. | +21° | *17* |
| 30 July 607 | 0.63 | 9.5 h. | +52° | *18* |
|  |  |  |  |  |
| 9 July 597 | 0.73 | 5.0 h. | +3 | 123 |
| 23 Dec 596 | 0.61 | 16.7 h. | 0° | *18* |
| 9 May 594 | 0.46 | 8.3 h. | +36° | *17* |
|  |  |  |  |  |
| 29 July 588 | 0.88 | 19.0 h. | +1° | 77 |
| 14 Dec 587 | 0.75 | 10.9 h. | +28 | *17* |
| 28 May 585 | 0.97 | 17.9 h. | +13 | *18* |

**Table 1**: Dirk Couprie's 'Table 4' with the alleged clusters of solar eclipses that Thales could have observed at Miletus.

| Date of solar eclipse | Maximum phase | Local time at maximum phase | Altitude of sun at maximum phase | Lunations elapsed since last solar eclipse |
|---|---|---|---|---|
| 17 June 633 | 0.66 | 19.2 h. | +2º | 29 |
| 17 April 611 | 0.45 | 18.6 h. | 0º | 270 |
| 30 Sept 610 | 0.59 | 8.6 h. | +30º | 18 |
| 13 Feb 608 | 0.76 | 15.2 h. | +21º | 17 |
| 30 July 607 | 0.63 | 9.5 h. | +52º | 18 |
| 18 May 603 | 0.50 | 8.2 h. | +36º | 47 |
| 9 July 597 | 0.73 | 5.0 h. | +3 | 76 |
| 23 Dec 596 | 0.61 | 16.7 h. | 0º | 18 |
| 9 May 594 | 0.46 | 8.3 h. | +36º | 17 |
| 29 July 588 | 0.88 | 19.0 h. | +1º | 77 |
| 14 Dec 587 | 0.75 | 10.9 h. | +28 | 17 |
| 28 May 585 | 0.97 | 17.9 h. | +13 | 18 |

**Table 2**: What Dirk Couprie's 'Table 4' would look like if we include all the potentially observable eclipses that match his visibility criterium.

|  | **Hartner notation for cycles** | **Hartner probability in the lunar case** | **Percentage in the solar case, mag ≥ 0.25** | **Percentage in the solar case, mag ≥ 0.50** | **Percentage in the solar case, mag ≥ 0.75** |
|---|---|---|---|---|---|
| 6 lunations | N* (ecl seasons) | 37.8 % | 4.6 % | 4.6 % | 2.3 % |
| 41 lunations | H* | 45.5 % | 3.8 % | 2.3 % | 2.3 % |
| 47 lunations | K | 46.3 % | 22.3 % | 18.4 % | 9.1 % |
| 88 lunations | B* | 51.4 % | 6.2 % | 1.1 % | 0.0 % |
| 129 lunations | C | 39.4 % | 12.3 % | 8.0 % | 4.5 % |
| 135 lunations | L* | 53.3 % | 3.8 % | 1.1 % | 0.0 % |
| 223 lunations | S (Saros) | 39.3 % | 10.8 % | 6.9 % | 2.3 % |
| 317 lunations | A | 42.3 % | 24.6 % | 19.5 % | 9.1 % |
| 358 lunations | M* | 42.7 % | 0.8 % | 1.1 % | 0.0 % |
| 446 lunations | D | 33.7 % | 8.5 % | 4.6 % | 0.0 % |
| 669 lunations | E (Exeligmos) | 76.2 % | 58.5 % | 49.4 % | 22.7 % |
| 804 lunations | F* | 52.7 % | 5.4 % | 1.1 % | 0.0 % |
| 939 lunations | G | 48.5 % | 34.6 % | 33.3 % | 22.7 % |
| 1074 lunations | T* | 60.6 % | 8.5 % | 2.3 % | 0.0 % |
| 1209 lunations | Q | 48.5 % | 20.8 % | 14.9 % | 6.8 % |

**Table 3**: Resulting percentages for our study on solar eclipse predictability assuming different magnitude limits (0.25, 0.50, 0.75), shown along with Hartner's probabilities for lunar eclipses and his notation for the cycles. Asterisks refer to eclipses occurring in alternate nodes.